\documentclass[twocolumn,showpacs,preprintnumbers,amsmath,amssymb,aps,pre]{revtex4-1}

\usepackage[dvips]{graphicx}% Include figure files
\usepackage{dcolumn}% Align table columns on decimal point
\usepackage{bm}% bold math
\usepackage[dvips]{epsfig}
\usepackage{ulem}  % To intoroduce strike \sout{text}
\usepackage{color} % To introduce colors  
\newcommand \be {\begin{equation}}
\newcommand \ee {\end{equation}}
\newcommand \bea {\begin{eqnarray}}
\newcommand \eea {\end{eqnarray}}

\begin{document}

\title{Creep motion of a model frictional system}
\author{Baptiste Blanc$^{1}$, Luis A. Pugnaloni$^{2}$ and Jean-Christophe G\'eminard$^{1}$}
\affiliation{$^{1}$Universit\'e de Lyon, Laboratoire de Physique, Ecole Normale Sup\'erieure de Lyon, CNRS, 46 All\'ee d'Italie, 69364 Lyon cedex 07, France.\\
$^{2}$Instituto de F\'{\i}sica de L\'{\i}quidos y Sistemas Biol\'{o}gicos (UNLP, CONICET La Plata), Calle 59 Nro. 789, 1900, La Plata, Argentina.}

\begin{abstract}
We report on the dynamics of a model frictional system
submitted to minute external perturbations. The system consists of 
a chain of sliders connected through elastic springs 
that rest on an incline.
By introducing cyclic expansions and contractions
of the springs we observe a reptation of the chain.
We account for the average reptation velocity theoretically.
The velocity of small systems exhibits a series of plateaus
as a function of the incline angle. Due to elastic effects,
there exists a critical amplitude below which the reptation
is expected to cease. 
However, rather than a full stop of the creep, we observe in numerical simulations a transition between a 
continuous-creep and an irregular-creep regime when the critical 
amplitude is approached.
The latter transition is reminiscent of the transition between
the continuous and the irregular compaction of granular matter
submitted to periodic temperature changes.

\end{abstract}

\maketitle

\section{Introduction.} 

Granular materials are collections of macroscopic particles (grains) that interact via dissipative
forces. As a result, external excitation is necessary to promote the motion of the grains.
In general, it is assumed that in absence of mechanical perturbations, such materials will eventually
reach a mechanically static configuration and remain at rest. In particular, thermal agitation will
not suffice to induce rearrangements and, for this reason, granular matter is said to be athermal \cite{knight95,philippe02,pouliquen03}.
This is however a simplified view that assumes an idealized situation in which all perturbation (temperature variations, humidity changes,
mechanical noise, etc.), even minute, can be suppressed.

In practice, small temperature variations can actually trigger small but measurable rearrangements of the structure.
Uncontrolled thermal dilations have been reported to generate stress fluctuations large enough to hinder reproducible measurements of the
stress field inside a granular pile \cite{vanel99,clement97}. 
They were even suspected to be the driving factor leading to large-scale 'static avalanches'
\cite{claudin97}. When accumulating, such minute perturbations can induce an irreversible 
evolution of the system. Several recent studies indeed showed that temperature cycles,
even of small amplitude, can induce the slow compaction (a creep) of dry granular materials
\cite{geminard03,chen06,chen09,divoux08,divoux09,divoux09t,divoux10}.
Moreover, a transition between a continuous-flow and an intermittent-flow regime, observed 
when the amplitude of the temperature cycles is decreased \cite{divoux08,divoux09,divoux09t,divoux10},
remains unexplained. Even if the transition is thought to be due to finite size effects ({\it i.e.} the finite
number of grains in the diameter of the tube), the mechanisms, in particular the role played by the confining walls,
are still under debate.

In the same manner, minute temperature changes have been identified, more than a century ago, to
induce the creep of solids in frictional contacts. H. Moseley reported first the descent, driven
by temperature variations, of lead plates covering the south side of the choir of Bristol cathedral
and mentionned that the same mechanism could be responsible for the motion of glaciers \cite{moseley}.
A macroscopic model, considering only the relative dilations of the solids in contact,
predicts that the creep velocity is proportional to the amplitude of the temperature changes
\cite{moseley,bouasse} and, thus, that any temperature variation, whatever its amplitude, leads to a motion.
However, as noticed by H. Bouasse, such a description does not take into account the elastic effects
which necessarily play a role for large systems \cite{bouasse}. When elastic effects come into play,
the temperature variations are sufficient to induce the motion only if their amplitude is larger
than a critical value which, in particular, increases with the size of the system \cite{croll09}.
Accordingly, for a given amplitude of the temperature variations, a solid is observed to descent
along an incline only if the tilt angle is large enough \cite{tamburi74}.

In the present article, we propose the study of a model mimicking the reptation along an incline
of a solid subjected to temperature variations. We remark that the aforementioned models disregard
the fact that the surfaces, even if nominally flat, are in real contact in a finite number of localized
regions. However, due to roughness, the real contact between the surfaces reduces to a large, but finite,
number of microcontacts, themselves belonging to a small number of mesoscopic, coherent, contact regions \cite{persson10}.
Such coherent regions are composed of a large number of microcontacts, so that their contact with the substrate obeys
the Amonton-Coulomb law for static friction. We thus consider a set of sliders, connected by springs, in frictional contact
with an incline and subjected to temperature changes. We report and discuss an extensive study of the associated dynamics.
We point out that, interestingly, the conclusions help in understanding the observations reported for a granular
material in a tube.

\section{The model}

\subsection{Description}

\begin{figure}[h!]
\includegraphics[width=0.9\columnwidth]{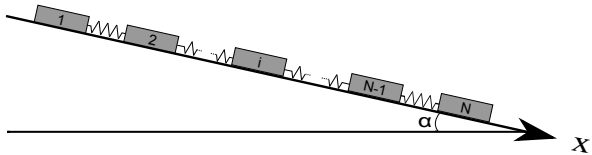}
\caption{{Sketch of the model system.}}
\label{sketch}
\end{figure}

The system (total mass $M$) consists of a series of $N$ identical sliders
(zero length, mass $m = M/N$) connected by $(N-1)$ identical springs (massless, stiffness $k$).
The sliders are in frictional contact with an incline which makes an angle, $\alpha$, with the horizontal (Fig.~\ref{sketch}).
They are subjected to the forces due to the springs, to their own weight $m g$
(where $g$ denotes the acceleration due to gravity) and to the reaction force from the incline
(including the frictional force).

The aim of the study is to account for the dynamics of the system induced by temperature changes.
To do so, we consider that the natural length of the springs, $l$, depends linearly on the temperature
$T$ according to:
\begin{equation}
l(T) = l_0 \bigl[ 1 + \kappa (T - T_0) \bigr]
\end{equation}
where $\kappa$ stands for the thermal expansion coefficient of the slider material and $l_0$ for the 
natural length of the springs at $T_0$. In accordance, the total length of the system at $T_0$,
in absence of internal stress, is $L = (N-1)\,l_0$.
We further assume, for the sake of simplicity, that the stiffness $k$ does not depend on the temperature
and that the substrate does not dilate. 

From now on, the $x$-axis is oriented downwards, $\alpha$ is positive
and the first slider is the upper one (Fig.~\ref{sketch}).
Thus, at the temperature $T$, the force due to the spring, exerted by the slider $(n+1)$ on the slider $n$ is:
\begin{equation} 
f_{n+1 \to n} = - k\,\left[ x_{n} - x_{n+1} + l(T) \right]
\label{spring_force}
\end{equation}
where $x_n$ denotes the position of the slider $n$ on the incline.

At rest, the contact between the sliders and the incline is characterized by the static
frictional coefficient $\mu_{\mathrm{s},n}$ such that the slider $n$, initially at rest,
starts moving if:
\begin{equation}
\left| f_{n+1 \to n} + f_{n-1 \to n} + m g\,\sin\alpha \right| > \mu_{\mathrm{s},n}\,m g\,\cos\alpha.
\label{static_friction}
\end{equation}
We remark that, at the boundaries, $f_{0 \to 1} =0$ ($n=1$) and $f_{N+1 \to N} =0$ ($n=N$).
Note that, due to the spatial heterogeneity of the incline surface,
the static frictional coefficient might take, at random, different values, $\mu_{\mathrm{s},n}$, 
for the different sliders which lie at different positions, $x_n$, on the incline \cite{baumberger06}.
In accordance, a slider that moves and stops at a different position might
be then associated with a different value of $\mu_{\mathrm{s},n}$.
By contrast, we assume that, when the slider is in motion, the frictional contact is characterized by
a single, constant, dynamical frictional coefficient, $\mu_\mathrm{d}$. 
Indeed, while in motion, the slider explores the incline and is thus less sensitive to the details of the surface properties.
The constant $\mu_\mathrm{d}$ quantifies the average rate of energy dissipation
and the slider $n$ in motion is subjected to the associated dynamical frictional force:
\begin{equation}
f_{\mathrm{d},n} \equiv - \mu_{\mathrm{d}}\,m g\,\cos\alpha\,S(\dot x_n)
\label{dynamical_friction}
\end{equation}
where $\dot x_n$ denotes the velocity and $S$ the sign function [$S(u)=1$ if $u>0$ and $S(u)=-1$ if $u<0$].
We further assume, in agreement with standard observations, that $\mu_\mathrm{d} < \mu_{\mathrm{s},n}$~$(\forall n)$,
i.e., the dynamical frictional coefficient is smaller than the static one.

The mechanical system is submitted to temperature changes.
Some sliders, initially at rest, start moving when the length $l(T)$ exceeds a 
value such that the condition \eqref{static_friction} is satisfied for, at least, one slider.
Submitted to the forces due to the springs, to their own weight and to the reaction force
from the incline (including the frictional force), one or more sliders move and come back to rest.
Such event constitutes, by definition, a ``{\it mechanical rearrangement}'' of the system.
At this point, it is interesting to consider two characteristic times.
On the one hand, due to its thermal inertia, the system exhibits a thermal characteristic time $\tau_\mathrm{th}$
which limits the dynamics of the temperature changes and, thus, of the thermal dilations.
In practice, for a macroscopic system whose typical size $L$ ranges from a few millimeters to centimeters,
$\tau_\mathrm{th}$ ranges from seconds to hours.
On the other hand, the dynamics of the mechanical system is characterized by the typical
time scale $\tau_\mathrm{dyn} = \sqrt{m/k}$. In practice, one can consider instead that
$\tau_\mathrm{dyn} \sim L/c_\mathrm{s}$ where $c_\mathrm{s}$ stands of the speed of sound
in the material the macroscopic solid is made of.
For a typical size $L$ ranging from a few millimeters to centimeters and usual values of $c_\mathrm{s}$
(about a few kilometers per second), we estimate $\tau_\mathrm{dyn} \sim 10^{-6} - 10^{-5}$~s,
thus much smaller than $\tau_\mathrm{th}$.
As a consequence, we consider that $l(T)$ is constant during the mechanical rearrangements
which are, in practice, much faster than the temperature variations.

\subsection{General system of equations}
\label{eq_system}

\subsubsection{Mechanical system}

First, we write the differential equation governing the position $x_\mathrm{n}$ of the slider $n$, when the latter
is in motion.
Introducing the thermal dilation $\theta \equiv \kappa\,(T - T_0)$ and the dimensionless
variables $\tilde{t} \equiv t/\tau_\mathrm{dyn}$ and $\tilde{x}\equiv x/(g\,\tau_\mathrm{dyn}^2)$ (with $\tau_\mathrm{dyn}\equiv\sqrt{m/k}$), we get:
\begin{eqnarray}
\ddot{\tilde{x}}_1 &=& - [\tilde{x}_1 - \tilde{x}_2 + (1 + \theta)\,\tilde{l}_0]\nonumber\\
&&- \mu_\mathrm{d}\,S(\dot{\tilde{x}}_1)\,\cos{\alpha}  + \sin\alpha~~~~~\left( n = 1 \right)\label{dynamic_eq1}\\
\ddot{\tilde{x}}_n &=& - [2\,\tilde{x}_n - ( \tilde{x}_{n+1} + \tilde{x}_{n-1})]\nonumber\\
&&- \mu_\mathrm{d}\,S(\dot{\tilde{x}}_n)\,\cos{\alpha}+ \sin\alpha~~~~~\left( n \neq 1, N\right)\label{dynamic_eq}\\
\ddot{\tilde{x}}_N &=& - [\tilde{x}_N - \tilde{x}_{N-1} - (1 + \theta)\,\tilde{l}_0]\nonumber\\
&&- \mu_\mathrm{d}\,S(\dot{\tilde{x}}_N)\,\cos{\alpha} + \sin\alpha~~~~~\left( n = N \right)\label{dynamic_eqN}
\end{eqnarray}
where, in accordance, we defined $\tilde{l}_0 \equiv l_0/(g\,\tau_\mathrm{dyn}^2)$.

Second, we remind that the slider $n$, if at rest, starts moving if the condition \eqref{static_friction}
is satisfied. Thus, using the dimensionless variables, we can write:
\begin{eqnarray}
| \tilde{x}_{2} -\tilde{x}_1 - (1 + \theta)\,\tilde{l}_0 + \sin\alpha | &>& \mu_{\mathrm{s},1}\,\cos\alpha \label{static_eq1}\\
| \tilde{x}_{n+1} + \tilde{x}_{n-1} -2\,\tilde{x}_n + \sin\alpha | &>& \mu_{\mathrm{s},n}\,\cos\alpha  \label{static_eq}\\
| \tilde{x}_{N-1} -\tilde{x}_N + (1 + \theta)\,\tilde{l}_0 + \sin\alpha | &>& \mu_{\mathrm{s},N}\,\cos\alpha \label{static_eqN}
\end{eqnarray} 
where the conditions \eqref{static_eq1} and \eqref{static_eqN} apply for the slider 1 and $N$ respectively
and the condition \eqref{static_eq} for any other slider $n \neq 1, N$. 

\subsubsection{Frictional contact}

We remind that the  dynamical frictional contact between the sliders and the incline is
characterized by a single coefficient, $\mu_\mathrm{d}$.
By contrast, each time a slider in motion comes back to rest, a new value of the static
frictional coefficient $\mu_\mathrm{s}$ is drawn from a probability distribution $p(\mu_\mathrm{s})$.
In order to evaluate if the fluctuations in $\mu_\mathrm{s}$ are relevant for explaining the
qualitative behaviour of the system and to be able to discuss the results analytically,
we assume that $p(\mu_\mathrm{s})$ is a Gaussian distribution, namely
\begin{equation}
p(\mu_\mathrm{s}) = \frac{1}{\sqrt{2\pi\sigma_\mu^2}} \exp\left[-\frac{( \mu_\mathrm{s} - \overline{\mu}_\mathrm{s})^2}{2 \sigma_\mu^2}\right]
\label{dist_mud}
\end{equation}
where  $\overline{\mu}_\mathrm{s}$ denotes the average value and $\sigma_\mu$ the width of the distribution.
We point out that, the distribution $p(\mu_\mathrm{s})$ does not, strictly speaking, satisfy
the condition $p(\mu_\mathrm{s})=0$, $\forall\mu_\mathrm{s}<\mu_\mathrm{d}$ (i.e., the static
friction coefficient can be smaller than the dynamic one).
However, we restrict our study to the case $\sigma_\mu \ll (\overline{\mu}_\mathrm{s}-\mu_\mathrm{d})$,
so that the latter condition is reasonably satisfied in practice.

\subsubsection{Temperature variations}
\label{temperature}

The system is driven by temperature variations which induce changes in the natural length of the springs.
We will either consider that the temperature oscillates with the period $2 \tau_\mathrm{th}$
between two well-defined values such that the dilation oscillates periodically between $-A_\theta$ and $+A_\theta$,
or that the temperature $T$ of the system fluctuates around the temperature $T_0$ with a typical 
amplitude $\Delta T$ such that $\theta$ fluctuates around zero with a typical amplitude $\sigma_{\theta} \sim \kappa\,\Delta T$. 
In this latter case, in order to be able to discuss the results analytically, we shall assume that,
at a series of equally-spaced timesteps $t_q \equiv q\,\tau_\mathrm{th}$ ($q \in \mathbb{N}$),
a value $\theta_q$ of the dilation is drawn from the Gaussian distribution:
\begin{equation}
\psi(\theta) = \frac{1}{\sqrt{2\pi\sigma_\theta^2}} \exp\left[-\frac{\theta^2}{2 \sigma_\theta^2}\right]
\label{dist_theta}
\end{equation}
where $\sigma_\theta$ accounts for the typical amplitude of the thermal dilations.
We then assume that $\theta = \theta_q$ for $t \in [t_q,t_{q+1}[$.

\subsection{Numerical method}
\label{numerics}

Consider that, when the time $t$ reaches $t_{q+1}$, the system already experienced $k$ mechanical rearrangements
such that all the sliders are resting at the positions $\tilde{x}_n^k$ 
associated with the static frictional coefficients $\mu_{\mathrm{s},n}^k$ ($n \in [1, N]$). 
At $t = t_{q+1}$, a new value $\theta_{q+1}$ of $\theta$ is drawn at random according to the distribution $\psi(\theta)$.
The subsequent evolution of the system is obtained as follows.

From the conditions \eqref{static_eq1}-\eqref{static_eqN}, one calculates the first critical value $\theta_c^{k+1}$
which leads to the destabilization of at least one slider (in general, $|\theta_c^{k+1}|<|\theta_{q+1}|$).
Then, for $\theta = \theta_c^{k+1}$, the equations \eqref{dynamic_eq1}-\eqref{dynamic_eqN}
are integrated numerically {using an integration time step $\Delta t\ll\tau_\mathrm{dyn}$}
(we use a standard velocity Verlet integrator \cite{numerical_recipes}).
In order to take into account that the motion of one slider can destabilize its neighbours,
the procedure checks if any condition \eqref{static_eq1}-\eqref{static_eqN} for the sliders at rest
is fulfilled within each timestep $\Delta t$.
If so, the motion of the corresponding slider(s) is also accounted for through \eqref{dynamic_eq1}-\eqref{dynamic_eqN}.
In the same way, the procedure checks at each timestep if any of the sliders in motion comes to rest.
If so, a new value of the associated static frictional coefficient is drawn at random according to the
distribution $p(\mu_\mathrm{s})$ [Eq.~(\ref{dist_mud})].
We consider that the {\it mechanical rearrangement} ends when all the sliders are back to rest.
All the sliders are then resting at the new positions $\tilde{x}_n^{k+1}$, associated with 
a new set of static frictional coefficients $\mu_{\mathrm{s},n}^{k+1}$.

The procedure is iterated by finding, in the same way, the next critical value
of the dilation $\theta_c$ which leads, according to the conditions
\eqref{static_eq1}-\eqref{static_eqN}, to the next mechanical rearrangement.
For each $\theta_c$, a new static state of the system is found.
The procedure stops when the next critical dilation $\theta_c$ is beyond $\theta_{q+1}$.
The system then experienced $k'$ rearrangements.
The $\tilde{x}_n^{k'}$ are then the positions of the sliders at $t = t_{q+2}$.

The long-term behavior of the system is assessed by iterating
the whole  procedure, either making $\theta$ oscillate between $-A_\theta$ and $+A_\theta$ or drawing at random the next value
of $\theta$ according to the distribution $\psi(\theta)$.

\section{Results}
\label{results}

\subsection{The minimal system: 2 sliders}
\label{two}

Consider first the system consisting of two sliders ($N=2$) connected by one spring.
The system is very similar to that already studied in Ref.~\cite{geminard10} for
the case of a horizontal substrate but, now, the system lies on an incline
which makes a finite angle $\alpha$ with the horizontal.

\subsubsection{Numerical results}

\begin{figure}[h!]
\noindent\includegraphics[width=\columnwidth]{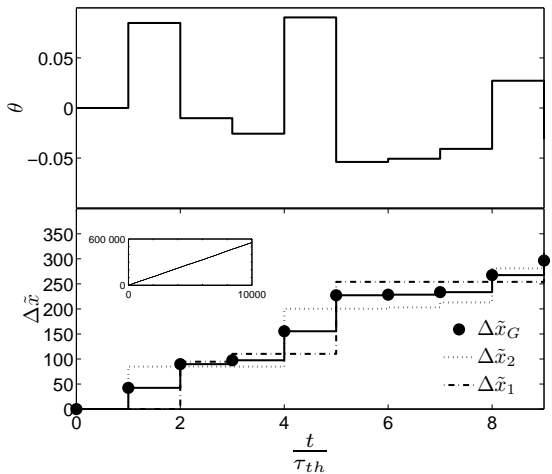}
\caption{\label{position2sliders}
{\bf Creep motion of 2 sliders -- }
Top: The dilation $\theta$ evolves randomly according to the Gaussian distribution (\ref{dist_theta}).
Bottom: In response to the dilations, the whole systems creeps along the incline. We report the displacements 
$\Delta\tilde{x}_1$ and $\Delta\tilde{x}_2$ of the sliders, as well as the displacement $\Delta\tilde{x}_G$ of the center of mass
with respect to their initial positions. We observe that
the lower slider moves in response to dilations whereas the upper slider moves in response to contractions.
Note that both sliders move systematically downwards. Inset: Long-time behavior -- $\Delta\tilde{x}_G$ increases
linearly with time $t$, the system is creeping at constant velocity 
($\tilde{l}_0 = 10^3$, $\mu_\mathrm{d}$=0.5, $\overline{\mu}_\mathrm{s}$=0.6, $\sigma_\mu$ = 0.01, $\sigma_\theta$ =0.1
and $\tan{\alpha}$ = 0.25).}
\end{figure}

In order to account for the creep of the system along the slope
we consider the position of the center of mass
$\tilde{x}_G \equiv \frac{1}{N}\,\sum_{n=1}^N \tilde{x}_n$ (here $N=2$)
as a function of time at the timesteps $t_q$ [Fig.~(\ref{position2sliders})]
for Gaussian variations of the temperature (Eq.~\ref{dist_theta}).
We observe that, for a sufficiently large $\sigma_\theta$,
the center of mass moves downwards with, in average, 
the constant velocity $<\tilde{v}_G>$.

\begin{figure}[h!]
\includegraphics[width=\columnwidth]{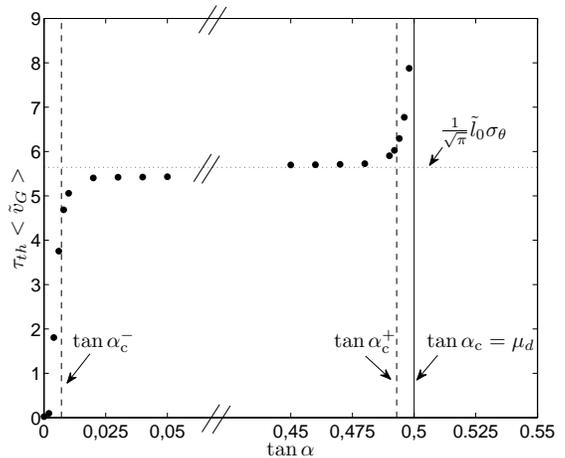}
\caption{{\bf Velocity $<\tilde{v}_G>$ vs. angle $\alpha$ --}
The average velocity of the center of mass,
$<\tilde{v}_G>$ depends only slightly on $\alpha$ (plateau) over a wide range of the incline angle typically
from $\alpha_\mathrm{c}^-$ to $\alpha_\mathrm{c}^+$. Below $\alpha_\mathrm{c}^-$, $<\tilde{v}_G>$ vanishes for vanishing $\alpha$.
Above $\alpha_\mathrm{c}^+$, $<\tilde{v}_G>$ drastically increases when $\alpha$ is increased and diverges for $\alpha = \alpha_\mathrm{c}$
($\tilde{l}_0 = 10^3$, $\mu_\mathrm{d}$=0.5, $\overline{\mu}_\mathrm{s}$=0.6, $\sigma_\mu$ = 0.01 and $\sigma_{\theta}$ = 0.01).}
\label{velocity2sliders}
\end{figure}

For a given amplitude of the temperature variations $\sigma_\theta$, we observe that the average creep velocity of
the center of mass, $<\tilde{v}_G>$, depends only slightly on the slope over a wide range of the angle $\alpha$ (Fig.~\ref{velocity2sliders}).
Indeed, for small $\alpha$, $<\tilde{v}_G>$, which equals zero for $\alpha=0$,
increases rapidly with $\alpha$ and reaches a plateau for a typical value that we denote $\alpha_\mathrm{c}^-$.
Then, above $\alpha_\mathrm{c}^-$, over a wide range of $\alpha$, $<\tilde{v}_G>$ increases only slightly
with $\alpha$ (plateau) until a second critical value $\alpha_\mathrm{c}^+$ is reached.
Above $\alpha_\mathrm{c}^+$, $<\tilde{v}_G>$ drastically increases as $\alpha$ approches the critical
angle of avalanche, $\alpha_\mathrm{c}$, defined by $\tan(\alpha_\mathrm{c}) \equiv \mu_\mathrm{d}$.

\subsubsection{Analysis}

The results presented above can be qualitatively understood by simple arguments.

First, for $\alpha=0$, by symmetry, we expect $<\tilde{v}_G> = 0$.
This case has been extensively studied in Ref.~\cite{geminard10}.
The system {\it ages} and the center of mass experiences a sub-diffusive motion.
Second, for $\alpha>\alpha_\mathrm{c}=\arctan(\mu_d)$, if the system starts sliding as a whole,
the $x$-component of the weight exceeds the dynamic friction force and the system accelerates. A finite average
velocity is not defined in this case.

The most striking result, {\it i.e.}, the plateau observed over a wide
range of the incline angle ($\alpha_\mathrm{c}^- \lesssim \alpha \lesssim \alpha_\mathrm{c}^+$),
can be easily understood by considering the mechanical stability of each of the sliders.
Indeed, assuming that the values of the static friction coefficients $\mu_{\mathrm{s},1}$ and
$\mu_{\mathrm{s},2}$ are close to each other, one can see that, because of the
force $m g\,\sin(\alpha)$ due to the gravity, the lower (resp. upper) slider moves
downwards when the system dilates (resp. contracts).  
As a consequence, if $\theta_{q+1} > \theta_{q}$ (the system dilates by $\Delta \theta \equiv \theta_{q+1} - \theta_{q} > 0$),
the resulting displacement of the center of mass is
$\Delta\tilde{x}_G = \frac{1}{2} \Delta\tilde{x}_2 \simeq \frac{1}{2}\tilde{l}_0\Delta\theta$. 
In the same way, if $\theta_{q+1} < \theta_{q}$ (the system contracts by $\Delta \theta < 0$),
$\Delta\tilde{x}_G = \frac{1}{2} \Delta\tilde{x}_1 \simeq -\frac{1}{2}\tilde{l}_0\Delta\theta$.
Thus, whatever the sign of $\Delta\theta$ (a contraction or a dilation),
$\Delta\tilde{x}_G \simeq \frac{1}{2}\tilde{l}_0 |\Delta\theta|$.
Note that $\Delta\tilde{x}_G$ does not depend on the angle $\alpha$.
Thus, when the system is subjected to periodic dilations from $\pm A_\theta$ to $\mp A_\theta$, the center of mass is displaced by
$\Delta\tilde{x}_G \simeq \frac{1}{2} \tilde{l}_0 (2 A_\theta)$ during ${\tau_\mathrm{th}}$
such that the average velocity of the system is given by:
\be
\tau_\mathrm{th} <\tilde{v}_G>\,\simeq \tilde{l}_0 A_\theta.
\label{velocity_th_2_cycle}
\ee
In the case of Gaussian fluctuations [Eq.~(\ref{dist_theta})], taking into account the distribution $\psi(\theta)$, we estimate 
the average velocity of the center of mass for $\alpha_\mathrm{c}^- \lesssim \alpha \lesssim \alpha_\mathrm{c}^+$:
\be
\tau_\mathrm{th} <\tilde{v}_G>\,\approx \frac{1}{\sqrt{\pi}} \tilde{l}_0 \sigma_\theta.
\label{velocity_th_2_gaussian}
\ee
The latter theoretical predictions (\ref{velocity_th_2_cycle}) and (\ref{velocity_th_2_gaussian})
are in good agreement with the numerical results
for large values of $A_\theta$ and $\sigma_\theta$
in a finite range of $\alpha$ (Fig.~\ref{velocity2sliders}).

Let us focus on the critical angles $\alpha_\mathrm{c}^-$ and $\alpha_\mathrm{c}^+$
which limit the plateau. On the one hand, the plateau velocity is reached if the angle $\alpha$ is large enough
($\alpha \gtrsim \alpha_\mathrm{c}^-$) for the force due to the gravity
to insure that the upper slider remains stable when the system dilates
and, conversely, that the lower slider remains stable when
the system contracts. For instance, for a dilation, the lower slider
(the slider 2) destabilizes first if
$f_{1\to2} + m g\,\sin\alpha > \mu_{\mathrm{s},2}\,m g\,\cos\alpha$ [Eq.~(\ref{static_eqN})]
and $ f_{2\to1} + m g\,\sin\alpha > - \mu_{\mathrm{s},1}\,m g\,\cos\alpha$ [Eq.~(\ref{static_eq1})]. 
The condition is thus, by summing the two inequalities,
$\tan\alpha > \frac{1}{2} (\mu_{\mathrm{s},2}-\mu_{\mathrm{s},1})$.
Taking into account the probability distribution $p(\mu_{\mathrm{s}})$,
we deduce that the latter condition is generally fulfilled if
$\tan\alpha > \tan\alpha_\mathrm{c}^- \equiv \sigma_\mu/\sqrt{2}$,
which defines $\alpha_\mathrm{c}^-$ (Fig.~\ref{velocity2sliders}).

On the other hand, the velocity of the system is larger than the plateau velocity if 
the motion of one slider can induce the motion of the second one ($\alpha \gtrsim \alpha_\mathrm{c}^+$).
The angle $\alpha_\mathrm{c}^+$ can be estimated by considering,
for instance, a dilation such that the slider 2 starts moving
for $f_{1\to2} + m g\,\sin\alpha > \mu_{\mathrm{s},2}\,m g\,\cos\alpha$ [Eq.~(\ref{static_eqN})].
Provided that the slider 1 remains at rest, the force $f_{1\to2}$ decreases
by $\Delta f_{1\to2} = 2\,m g\,\cos\alpha\,(\mu_{\mathrm{s},2}-\mu_\mathrm{d})$ [by solving the equation~(\ref{dynamic_eqN})].
The slider 1 remains stable if $-f_{1\to2}+m g\,\sin\alpha < \mu_{\mathrm{s},1}\,m g\,\cos\alpha$ [Eq.~(\ref{static_eq1})],
i.e. if $\tan\alpha < \mu_\mathrm{d} + {(\mu_{\mathrm{s},1}-\mu_{\mathrm{s},2})}/{2}$.
We obtain a similar result by considering a contraction of the system. 
Thus, considering the width of the distribution $p(\mu_\mathrm{s})$ [Eq.~(\ref{dist_mud})],
we estimate that the second slider generally remains stable provided that
$\tan\alpha < \tan\alpha_\mathrm{c}^+ \equiv \mu_\mathrm{d} - \sigma_\mu/{\sqrt{2}}$,
which defines $\alpha_\mathrm{c}^+$ (Fig.~\ref{velocity2sliders}).

\begin{figure}[h!]
\includegraphics[width=\columnwidth]{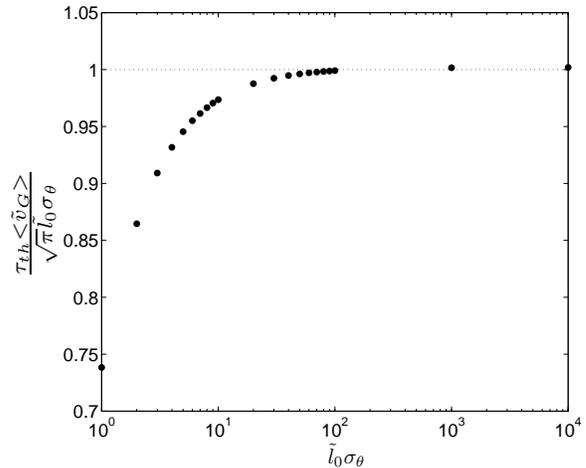}
\caption{{\bf Velocity $<\tilde{v}_G>$ vs. typical amplitude $\sigma_\theta$ --}
In accordance with equation (\ref{velocity_th_2_gaussienne_corrected}), the average velocity $<\tilde{v}_G>$
tends to the asymptote ${\tilde{l}_0} \sigma_\theta / (\sqrt{\pi} \tau_\mathrm{th})$ only in the limit $\sigma_\theta \gg \sigma_\theta^*$
(Here $\tilde{l}_0 \sigma_\theta^*$ is about 0.4 whereas the asymptote is reached to within 1\% for $\tilde{l}_0 \sigma_\theta$ about 100,
$\tilde{l}_0 = 10^3$, $\mu_\mathrm{d}$=0.5, $\overline{\mu}_\mathrm{s}$=0.6, $\sigma_\mu$ = 0.01 and $\tan{\alpha}$ = 0.25).}
\label{velocity2sliders_theta}
\end{figure}
The simplistic reasoning proposed above leads to the prediction of a 
well-defined plateau between $\alpha_\mathrm{c}^-$ and $\alpha_\mathrm{c}^+$.
However, a careful analysis of the numerical data reveals that $<\tilde{v}_G>$ slightly
increases with the angle $\alpha$ even for large amplitudes $A_\theta$ or $\sigma_\theta$
(Fig.~\ref{velocity2sliders}).
To obtain the plateau values [Eq.~(\ref{velocity_th_2_cycle})] or [Eq.~(\ref{velocity_th_2_gaussian})],
we implicitly assumed that  any variation of $\theta$ was enough for one of the conditions
(\ref{static_eq1}) and
(\ref{static_eqN}) to be fulfilled, which is not correct
when a dilation is followed by a contraction (or conversely).
In this case, taking into account that the stable 
and unstable sliders are swapped in Eqs.~(\ref{static_eq1}) and
(\ref{static_eqN}), we obtain
$\Delta\tilde{x}_G \simeq \frac{1}{2} \Bigl[ \tilde{l}_0 |\Delta\theta|
- 2 (\overline{\mu}_\mathrm{s} \cos{\alpha} - \sin{\alpha}) \Bigr]$.
The sliding distance is reduced by the fact that a minimum dilation is 
necessary to invert the direction of the frictional force.
As a consequence, in the case of periodic cycling, the velocity is given by:
\be
\tau_\mathrm{th} <\tilde{v}_G>\,\simeq \tilde{l}_0 A_\theta - (\overline{\mu}_\mathrm{s} \cos{\alpha} - \sin{\alpha}),
\label{velocity_th_2_cycle_corrected}
\ee
which explains the mild increase of $<\tilde{v}_G>$ with $\alpha$ for $\tilde{l}_0 A_\theta \gg 1$. 
The relation (\ref{velocity_th_2_cycle_corrected}) holds valid provided that
$A_\theta > A_\theta^*$ defined by:
\be
A_\theta^* \equiv \frac{1}{\tilde{l}_0} (\overline{\mu}_\mathrm{s} \cos{\alpha} - \sin{\alpha}).
\label{threshold_th_2_cycle_corrected}
\ee
If the amplitude is smaller than $A_\theta^*$, the dilations are not sufficient to rearrange the system
and $<\tilde{v}_G> = 0$.
In the case of Gaussian variations [Eq.~(\ref{dist_theta})], a dilation (resp. contraction)
is followed by a contraction (resp. dilation) in 2/3 of the timesteps.
As a consequence, in a first approximation, the average velocity is given by
\be
\tau_\mathrm{th} <\tilde{v}_G>\,\approx \frac{\tilde{l}_0}{\sqrt{\pi}} \sigma_\theta
- \frac{2}{3} (\overline{\mu}_\mathrm{s} \cos{\alpha} - \sin{\alpha}).
\label{velocity_th_2_gaussienne_corrected}
\ee
where we assumed that all dilations are large enough to rearrange the system in each of the timesteps.
The condition is reasonnably fulfilled when $\sigma_\theta$ is much larger than:
\be
\sigma_\theta^* \equiv \frac{2\sqrt{\pi}}{3\tilde{l}_0} (\overline{\mu}_\mathrm{s} \cos{\alpha} - \sin{\alpha}).
\label{threshold_th_2_gaussienne_corrected}
\ee
Thus, for $\sigma_\theta \gg \sigma_\theta^*$, $<\tilde{v}_G>$ increases linearly with $\sigma_\theta$
[Eq.~(\ref{velocity_th_2_gaussienne_corrected})]. When $\sigma_\theta$ is deacreased, even below $\sigma_\theta^*$,
the dilations are always likely to rearrange the system and $<\tilde{v}_G>$ continuously decreases
and vanishes only for $\sigma_\theta = 0$ (Fig.~\ref{velocity2sliders_theta}). 

\subsection{Large systems: more than 2 sliders}
\label{anyN}

Consider now a system consisting of N sliders connected by N-1 springs. 
It is important to notice that, for $N > 2$, the system differs qualitatively 
from the system made of 2 sliders. Indeed, for $N > 2$, the dilation can induce
the motion of the sliders at both ends without, necessarily inducing, a displacement
of the slider(s) at center. Thus, the successive dilations do not necessarily induce
a displacement of the center of mass in average.

\subsubsection{Numerical results}
\label{numericsN}

\begin{figure}[h!]
\includegraphics[width=\columnwidth]{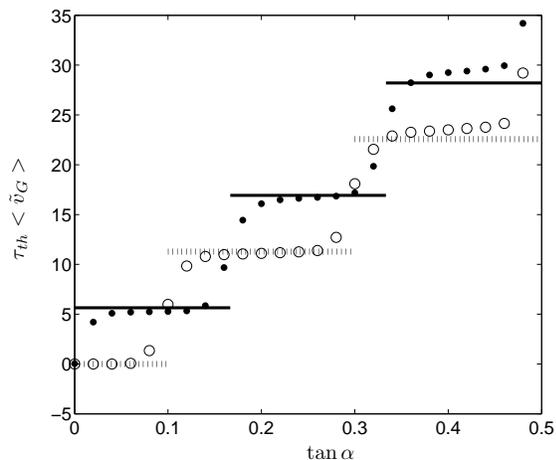}
\caption{{\bf Velocity $<\tilde{v}_G>$ vs. incline angle $\alpha$ --}
The velocity $<\tilde{v}_G>$ exhibits a series of plateaus corresponding to constant velocities $v_\mathrm{N,i}^*$
above critical angles $\alpha_\mathrm{N,i}^*$. The number of plateaus increases with the number of sliders
[dotted lines and open circles: $N=5$; full lines, and full circles: $N=6$. The plateau values (lines) are from Eq.~(\ref{vitesse})].
The velocity $<\tilde{v}_G>$ diverges when $\alpha$ approaches the critical angle $\alpha_\mathrm{c}$
($\tilde{l}_0 = 10^3$, $\mu_\mathrm{d}$=0.5, $\overline{\mu}_\mathrm{s}$=0.6,
$\sigma_\mu$ = 0.01 and $\sigma_\theta$ =0.01).}
\label{velocityNsliders}
\end{figure}

In order to account for the creep of the system along the slope
we consider the position of the center of mass $\tilde{x}_G \equiv \frac{1}{N}\,\sum_{n=1}^N \tilde{x}_n$
as a function of time at the timesteps $t_q$ and report the average velocity $<\tilde{v}_G>$,
for large amplitudes $A_\theta$ or $\sigma_\theta$, as a function of the incline angle $\alpha$ (Fig.~\ref{velocityNsliders}).
We observe that, for a small number $N$ of sliders, $<\tilde{v}_G>$ exhibits a series 
of plateaus: $<\tilde{v}_G>$ takes an almost constant value
in a finite range of the incline angle $\alpha$.
The number of plateaus increases when the number $N$ of sliders is increased.
As expected, $<\tilde{v}_G>$ drastically increases when $\alpha$ approaches the critical angle $\alpha_\mathrm{c}$. 

In the next section, we estimate theoretically the set of critical angles and the values of the corresponding
plateau velocities in the large amplitude limit.
In addition, we consider the dependence of the velocity on the amplitude of the temperature variations.

\subsubsection{Analytic estimates}
\label{theoryN}

The behavior of the system can be understood by considering the 
motion of the internal sliders. Let us assume that, during a dilation of large amplitude,
a given slider $i$ does not move whereas the sliders below move
downwards and the sliders above move upwards.
In the same way, let us assume that during a large contraction, a given slider $j$ does 
not move whereas the sliders above move downwards and 
the sliders below move upwards. If the incline angle $\alpha$ is large
enough, the sliders $i$ and $j$ differ (with $i<j$). In this case,
the internal displacements result in a {\it reptation} of the entire system along the incline
as already described in the framework of the continuous models \cite{moseley,bouasse,tamburi74}.

Let us first determine the critical angle $\alpha_\mathrm{N,i}^*$ above which
the slider $i$, previously at rest, starts moving downwards during a dilation,
the slider $i-1$ remaining at rest instead. Thus, let us first consider that
the slider $i$ remains at rest. We assume, in a first approach, that the dilation 
rate is large enough for the sliders in motion to slide continuously such that
they are submitted to the dynamical frictional force $\pm \mu_\mathrm{d} m g \cos \alpha$.
The condition is fulfilled provided that $\dot \theta \gg \Delta\mu / \tilde{l}_0 \tau_\mathrm{dyn}$
where we define $\Delta\mu \equiv \mu_\mathrm{s} - \mu_\mathrm{d}$.
In addition, the dilation must be large enough for all the downstream sliders to move
downwards and all the upstream sliders to move upwards.
We shall later discuss this assumption. In this case, neglecting the inertia,
one can indeed write the forces exerted by the slider $i$ on the sliders $i-1$ and $i+1$:
\begin{eqnarray}
&f_{i\to i-1} + (i-1) &(m g \sin \alpha + \mu_\mathrm{d} m g \cos \alpha) = 0\nonumber\\
&f_{i\to i+1} + (N-i) &(m g \sin \alpha - \mu_\mathrm{d} m g \cos \alpha) = 0\nonumber
\end{eqnarray}
Considering the condition for the stability of the slider $i$,
\be
| f_{i-1\to i} + f_{i+1\to i} + m g \sin \alpha | < \mu_\mathrm{d} m g \cos \alpha,\nonumber
\ee
we obtain that the slider $i$ starts moving downwards above the critical angle $\alpha_\mathrm{N,i}^*$
given by:
\be
\tan \alpha_{N,i}^* = \mu_\mathrm{d} \Bigl[ 1 - 2\,\frac{i-1}{N} \Bigr] + \frac{\Delta\mu}{N}.
\label{critical_angle}
\ee
Note that $\alpha_\mathrm{N,i}^*$ is a decreasing function of $i$ such that the slider $i-1$
remains stable for $\alpha_\mathrm{N,i}^* < \alpha < \alpha_\mathrm{N,i-1}^*$.

Let us now consider a contraction of the same system.
Assuming that for the chosen value of $\alpha$, the slider $i$ is at rest during the dilation,
considering that the slider $j$ remains at rest during the contraction, we write:
\begin{eqnarray}
&f_{j\to j-1} + (j-1) &(m g \sin \alpha - \mu_\mathrm{d} m g \cos \alpha) = 0\nonumber\\
&f_{j\to j+1} + (N-j) &(m g \sin \alpha + \mu_\mathrm{d} m g \cos \alpha) = 0\nonumber
\end{eqnarray}
Considering the stability of the slider $j$, replacing $\alpha$ by the critical value $\alpha_\mathrm{N,i}^*$,
we obtain that for $\alpha > \alpha_{N,i}^*$ the slider $j = N + 1 -i$ starts moving downwards such that the
slider $j+1$ is then at rest.

In summary, for $\alpha_\mathrm{N,i}^* < \alpha < \alpha_\mathrm{N,i-1}^*$, the slider $i-1$ remains stable
during the dilation and the slider $j+1$ remains stable during the contraction of the system.
Accordingly, we can write the displacements of any slider $n$, for a dilation
$\Delta\tilde{x}_\mathrm{n} = [ n - ( i - 1 ) ]\,\tilde{l}_0  \Delta\theta^+$ ($\Delta\theta^+ > 0$),
for a contraction 
$\Delta\tilde{x}_\mathrm{n} = [ n - ( j + 1 ) ]\,\tilde{l}_0  \Delta\theta^-$ ($\Delta\theta^- < 0$).
For a cycle $\Delta \theta^+ = - \Delta \theta^- = 2 A_\theta$,
one obtains the total displacement  $\Delta\tilde{x}_\mathrm{n} = 2 (N + 3 - 2 i) \tilde{l}_0 A_\theta$
for a total duration $2\,\tau_\mathrm{th}$. Note that $\Delta \tilde{x}_\mathrm{n}$ does not
depend on $n$ and that the associated velocity of the center of mass is simply
\be
\tau_\mathrm{th} v_\mathrm{N,i}^* =  ( N + 3 - 2 i )\,\tilde{l}_0  A_\theta.
\label{vitesse}
\ee
We thus expect, for cycles of amplitude $A_\theta$, the plateau velocity
$v_\mathrm{N,i}^*$ for $\alpha_\mathrm{N,i}^* < \alpha < \alpha_\mathrm{N,i-1}^*$ and $1 < i < (N+3)/2$
(note that, for $i=1$, $\alpha_\mathrm{N,1}^*$ corresponds to the critical angle of avalanche $\alpha_c$).
For Gaussian temperature variations $A_\theta$ must be replaced by $\sigma_\theta/\sqrt{\pi}$ so that 
$\tau_\mathrm{th} v_\mathrm{N,i}^* =  ( N + 3 - 2 i )\,\tilde{l}_0  \sigma_\theta/\sqrt{\pi}$
in this case.
We observe in Fig.~\ref{velocityNsliders} that the equations (\ref{critical_angle}) and (\ref{vitesse})
give good estimates of the transitions and plateau velocities.

\begin{figure}[h!]
\includegraphics[width=\columnwidth]{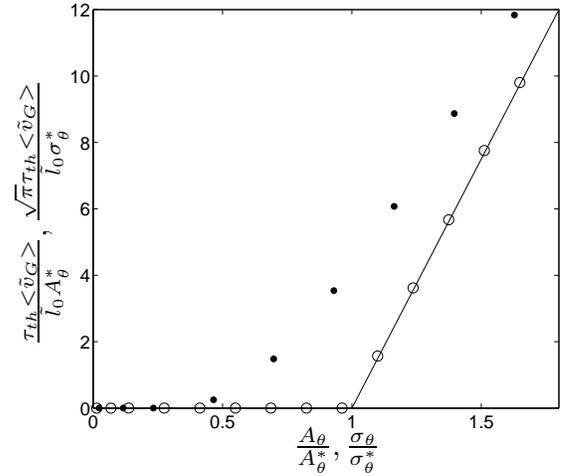}
\caption{{\bf Velocity $<\tilde{v}_G>$ vs. amplitude $A_\theta$ or $\sigma_\theta$ --}
For temperature cycles (open circles), $<\tilde{v}_{g}>$ increases linearly with the amplitude $A_\theta$ above a critical amplitude 
$A_\theta^*$ in agreement with the equation (\ref{vitesse_seuil}) [continuous line].
For random temperature variations (full circles), $<\tilde{v}_{g}>$ increases significantly above a critical amplitude
$\sigma_\theta^*$ of the temperature variations and reaches a linear asymptote for $\sigma_\theta \gg \sigma_\theta^*$
in agreement with the equation (\ref{vitesse_seuil_random}) [continuous line. We remind that the asymptote is reached only
when $\sigma_\theta \gg \sigma_\theta^*$].
Random temperature variations thus lead to a smoother transition but do not change the qualitative behavior of the system
($N = 30$, $\tilde{l}_0 = 10^3$, $\mu_\mathrm{d}$=0.5, $\mu_\mathrm{s}$=0.6, $\sigma_\mu$ = 0.01, and $\tan\alpha=0.25$).}
\label{velocityNslidersbis}
\end{figure}

When a contraction follows a dilation (or conversely), a compression (resp. dilation) wave propagates from
both ends inwards. Equation (\ref{vitesse}) is correct provided that the amplitude $A_\theta$ is larger
than a critical amplitude $A_\theta^*$, such that the temperature variations induce the motion
of all the sliders in the chain, which can be assessed in the following way:
For $\alpha_\mathrm{N,i}^* < \alpha < \alpha_\mathrm{N,i-1}^*$ and $A_\theta > A_\theta^*$,
during a dilation, the slider $i-1$ pushes all the sliders below downwards so that:
\be
f_{i-1 \to i} + [N - (i-1)] (\sin \alpha - \mu_\mathrm{d} \cos \alpha) = 0
\label{forcestatique}
\ee
Note that during the dilation the sliders $i$ to $j$ are moving downwards so that the frictional
force is already oriented upwards. During the contraction that follows, the sliders $1$ to $j$
move downwards provided that the contraction wave propagating from the upper end reaches the slider
$i-1$. Let us now denote $\Delta \theta^*$ the corresponding dilation. For $\Delta \theta^*$, 
the slider $i-1$, which is still at rest, pulls all the sliders above downwards so that:
\be
f_{i-1 \to i-2} + (i-2) (\sin \alpha - \mu_\mathrm{d} \cos \alpha) = 0\nonumber.
\ee
The sliders $i-1$ and $i$ being still at rest, we get from equation (\ref{forcestatique}),
\be
f_{i \to i-1} = [N - (i-1)] (\sin \alpha - \mu_\mathrm{d} \cos \alpha) - \tilde{l}_0 \Delta\theta^* \nonumber
\ee
The temperature variation induces the motion of the entire chain provided that
\be
f_{i \to i-1} + f_{i-2 \to i-1} + \sin \alpha - \mu_\mathrm{s} \cos \alpha > 0\nonumber
\ee
which leads to $- \tilde{l}_0 \Delta\theta^* > N (\sin \alpha - \mu_\mathrm{d} \cos \alpha) - \Delta\mu \cos \alpha$
(remember here that $\Delta\theta^* < 0$). Within a cycle, the amplitude of a contraction being
$\Delta\theta = -2 A_\theta$ and the argument developped above holding true for a dilation following
a contraction, we obtain the minimal amplitude leading to the reptation of the chain:
\be
A_\theta^* = \frac{N}{2\tilde{l}_0} (\mu_\mathrm{d} \cos \alpha - \sin \alpha)
\label{seuil}
\ee
where we assumed $\Delta\mu / N \ll 1$. 
Note that $A_\theta^*$ is independent of the position of the steady slider ($i$ or $j$)
and proportional to the number $N$ of sliders in the chain. We can thus write, for finite amplitudes,
\be
\tau_\mathrm{th} v_\mathrm{N,i}^* =  ( N + 3 - 2 i )\,\tilde{l}_0  \Bigl[ A_\theta -
\frac{N}{2\tilde{l}_0} (\mu_\mathrm{d} \cos \alpha - \sin \alpha) \Bigr].
\label{vitesse_seuil}
\ee
The equation (\ref{vitesse_seuil}) is in excellent agreement with the numerical results
even when the static and dynamical frictional coefficients are not equal ($\Delta \mu \neq 0$, Fig.~\ref{velocityNslidersbis}). 
We thus also deduce from these observations that the average velocity is not sensitive to the width 
$\sigma_\mu$ of the distribution of the static frictional coefficient.

In addition, the same qualitative behavior is expected, when the system is submitted to random temperature
variations. In Fig.~\ref{velocityNslidersbis}, we observe that equation~(\ref{vitesse_seuil}) agrees
with the numerical results, provided that the amplitude $A_\theta$
is replaced by $\sigma_\theta/\sqrt{\pi}$ and $A_\theta^*$ by $\sigma_\theta^* = \frac{2}{3} \sqrt{\pi} A_\theta^*$ so that:
\be
\tau_\mathrm{th}v_\mathrm{N,i}^* =  (N+3-2i)\,\tilde{l}_0 \Bigl[\frac{\sigma_\theta}{\sqrt{\pi}} - 
\frac{N}{3\tilde{l}_0} (\mu_\mathrm{d}\cos\alpha-\sin\alpha)\Bigr]
\label{vitesse_seuil_random}
\ee
for random temperature variations.

\section{Discussion}
\label{discussion}
\begin{figure}[t!]
\includegraphics[width=\columnwidth]{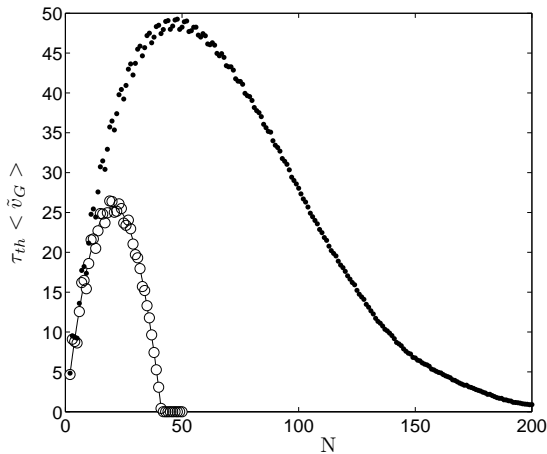}
\caption{{\bf Velocity $<\tilde{v}_G>$ vs. number of sliders $N$ --}
For temperature cycles (open circles) and random temperature variations (full circles),
$<\tilde{v}_{g}>$ exhibits a non-monotonic behavior as a function of $N$.
For small $N$, $<\tilde{v}_{g}>$ increases because of the increase of the size of the system; 
for large $N$, $<\tilde{v}_{g}>$ decreases because of the increase of critical amplitude $A_\theta^*$ or $\sigma_\theta^*$.
For temperature cycles, the numerical results are in excellent agreement with the equation (\ref{vitesse_seuil}) (continuous line).
Note however the dispersion of the numerical points for small $N$ which is due to the fact that, at a given amplitude $A_\theta$, 
$<\tilde{v}_{g}>$ corresponds to discrete plateau values which are not accounted for by the equation (\ref{vitesse_seuil})
when considering continuous values of $\alpha$
($\tilde{l}_0 = 10^3$, $\mu_\mathrm{d}$=0.5, $\overline\mu_\mathrm{s}$=0.6, $\sigma_\mu$ = 0.01, $A_\theta$ = 0.005
or $\sigma_\theta = 0.005 \sqrt{\pi}$ and $\tan \alpha=0.25$).}
\label{cloche}
\end{figure}

In the continuous description proposed by Moseley and Bouasse \cite{moseley,bouasse}, the elasticity
of the material was neglected which led to the conclusion that the creep velocity was proportional 
to the amplitude of the temperature variations. The introduction of the elastic effects leads to 
the conclusion that the temperature variations induce the creep of the system only if their amplitude
is large enough, as already  proposed by Croll. However, note that the critical value given by the
equation (\ref{seuil}) differs from the result proposed in \cite{croll09}.
The difference comes from the fact that Croll considered
that the system was free of stress at the beginning of each phase of a cycle (dilation or contraction).
In our approach, each dilation (resp. contraction) follows a contraction (resp. dilation) and the
frictional force is initially non zero, mobilized in the opposite direction.  
Considering that our chain of sliders model a solid of mass $M$, length $L$, cross section $S$
made of a material having a Young modulus $Y$ and a density $\rho$ creeping along an incline,
one can estimate from equation (\ref{vitesse}), provided that $k = N Y S / L$,
\be
\tau_\mathrm{th} v_G =
L \frac{\tan \alpha}{\mu_\mathrm{d}}
\Bigl[A_\theta - \frac{g \rho L}{2 Y} (\mu_\mathrm{d} \cos \alpha - \sin \alpha) \Bigr].
\label{vitessecontinue}
\ee
We thus obtain that the creep velocity is independent of the number $N$ of contacts.
The minimum amplitude of the temperature changes that produce the creep of
the system along the incline can be written:
\be
\Delta T^* = \frac{g \rho L}{2 Y \kappa} (\mu_\mathrm{d} \cos \alpha - \sin \alpha).
\ee
Thus, for a given $L$, the system is thus more likely to creep along the incline
for a larger Young modulus $Y$, a larger thermal expansion coefficient $\kappa$
and a larger angle $\alpha$. Note again that the stability of the system does not
depend on the number $N$ of contacts.
Typically, with $L = 1$~cm, $Y \sim 100$~GPa and $\rho \sim 10^4$~kg.m$^{-3}$,
we estimate that relative dilations of about $10^{-8}$ are enough to make
the system creep. Such dilations, which correspond to temperature changes of about 
$\Delta T \sim 1$~mK, are difficult to avoid and, in general, any frictional
contact between two macroscopic solids cannot be considered as perfectly static.

\begin{figure}[h!]
\includegraphics[width=\columnwidth]{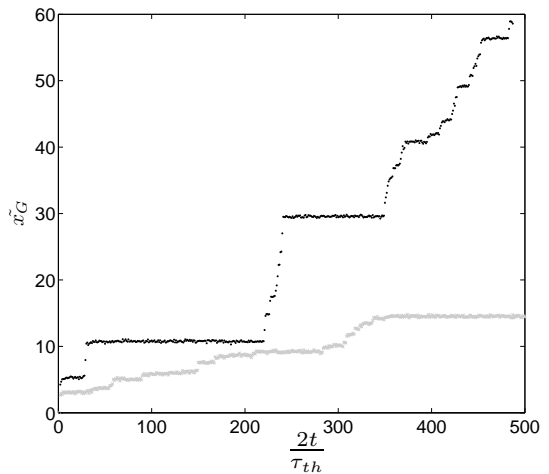}
\caption{{\bf Position $<\tilde{x}_G>$ vs. time ${t}$ --}
For $A_\theta \simeq A_\theta^*$, the system exhibits an irregular dynamics even when temperature cycles are imposed.
Note that fast creep can be followed by long quiescent periods as observed for $A_\theta = 1.01 A_\theta^*$ (black)
and $A_\theta = 0.99 A_\theta^*$ (grey) for $A_\theta^* \simeq 3.6~10^{-3}$ 
($N=30$, $\tilde{l}_0 = 10^3$, $\mu_\mathrm{d}$=0.5, $\mu_\mathrm{s}$=0.6, $\sigma_\mu$ = 0
and $\tan \alpha =0.25$).}
\label{jumps}
\end{figure}
The second practical situation mentionned in the introduction is the creep of granular
material induced by temperature changes \cite{geminard03,chen06,chen09,divoux08,divoux09,divoux09t,divoux10}. 
In this case, the length $l_0$ would account for the typical grain size, $k$ for the rigidity 
associated with the grain-grain contact and
$N$ for the typical number of grains in one typical dimension $L = N l_0$ of the system.
It is then particularly interesting to consider the dependence of the 
typical creep velocity on the number $N$ [Eq.~(\ref{vitesse_seuil}), Fig.~\ref{cloche}].
For a given amplitude of the temperature variations, $<\tilde{v}_{g}>$ exhibits a non-monotonic
behavior as a function of $N$, increasing linearly with $N$ for small $N$ and
decreasing for large $N$ because of the increase of the critical amplitude of the temperature
variations likely to induce the creep [Eq.~(\ref{seuil})]. For temperature cycles, from 
equation (\ref{seuil}), $<\tilde{v}_{g}>$ is expected to vanish for
$N > {2\tilde{l}_0} A_\theta / (\mu_\mathrm{d} \cos \alpha - \sin \alpha)$
whereas random temperature variations are expected to be always likely to produce creep.

Let us now focus on the dynamics of the system subjected to temperature cycles
of amplitude close to $A_\theta^*$ (Fig.~\ref{jumps}). We observe that, for 
$A_\theta$ close to $A_\theta^*$ (above and below $A_\theta^*$) the system creeps in an irregular manner.
The position of the center of mass, $\tilde{x}_\mathrm{G}$, exhibits a series of rapid
variations (jumps), separated by periods of time during which the system is apparently
at rest. This latter conclusion holds true even if the value of the static frictional coefficient
is well-defined, {\it i.e.} even for $\sigma_\mu =0$.
Thus, for $A_\theta$ approaching $A_\theta^*$, one observes a transition from 
a continuous creep regime, during which $\tilde{x}_\mathrm{G}$ decreases at each cycle,
to the irregular creep regime. The transition is reminiscent of the
transition between a continuous-flow and an intermittent-flow regime, observed 
when the amplitude of the temperature cycles imposed to a granular column is decreased
\cite{divoux08,divoux09,divoux09t,divoux10}. The experimental and theoretical systems
are very different but there are some common aspects like the frictional contact between the particles
that are in a limited number $N$. We can however attempt to estimate the critical 
amplitude $A_\theta^*$ expected from the model for a column (diameter 1 cm) of glass beads (typically 500 $\mu$m
in diameter). The elasticity of the system is due to the Hertz contact between the grains
and we can estimate that the stiffness $k$ depends on the pressure. Denoting $\delta$ the penetration
distance and $R$ the radius of the grains, we can write $k \sim Y (R\delta)^{1/2}$.
For an infinite vertical column, because of Janssen effect, one can estimate the local
pressure $P \sim \rho g D$, where $D$ stands for the diameter of the column and $\rho$
for the density of glass. Writing that the force applied to the grains $k \delta \sim P R^2$,
we get $k \sim R [\rho g D Y^2]^{1/3}$. From equation (\ref{seuil}), with $l_0 \sim 2R$ and
$N \sim D/2R$, we estimate $A_\theta^* \sim (\rho g D / Y)^{2/3}$. Note first that the
result does not depend on the size of the grains and, thus, not on the number of grains 
in the diameter of the column. Second, with $\rho \sim 2\,10^3$, $D \sim 1$~cm and $Y \sim 20$~GPa,
we obtain that $A_\theta^* \sim 5\,10^{-6}$ and, thus, with $\kappa \simeq 3\,10^{-6}$~K$^{-1}$
for glass, that a transition between the continuous- and the irregular-flow regimes is
expected for amplitudes of the temperature variations of the order of 1~K.
The latter value is interestingly close to the experimental value which
is of about 3~K \cite{divoux08}. 
Even if the good agreement between the theoretical estimate and the experimental values might be accidental,
we think that the comparison between our model and the granular column is worth
to be mentionned.

\section{Conclusion}
\label{conclusion}

We reported on the detailed behavior of a frictional system subjected to thermal dilations.
We observed that for a small number of sliders in the chain and large amplitude of the dilations,
the 'reptation' velocity of the center of mass exhibits a series of plateaus as a function of the
incline angle. In the limit of an infinite number of sliders and large amplitudes, we recover former
results obtained for the creep of a solid on an incline. Because of the elasticity of the material,
the creep velocity is expected to vanish for a finite value when the amplitude of the cycles is
decreased. We obtain an expression of the critical velocity which slighly differs from former results. 

Finally, for a finite number of sliders, we observe, numerically, that the system experiences
an irregular trajectory, the center of mass sliding rapidly between quiescent, rather long, 
periods of time even if the system is subjected to periodic cycling. The transition 
between the continuous and the irregular creep depends on the size of the system. 
The irregular creep is reminiscent of recent observations of the irregular compaction of
granular matter under the action of periodic temperature changes. 
The systems are very different but we do believe that the study of this specific
regime will provide us with interesting clues for the understanding of the peculiar behavior
of granular matter. The latter study will be the subject of a further publication.

\acknowledgments{}
The authors acknowledge financial support from
the {\it Agence Nationale de la Recherche} (contract ANR-09-BLAN-0389-01)
and from the CNRS/Conicet international cooperation action.

%\bibliographystyle{prsty}
%\bibliography{savedrecs}

%merlin.mbs 2010-03-15 4.21a (PWD, AO, DPC)
\begin{thebibliography}{0}%
\makeatletter
\providecommand \@ifxundefined [1]{%
 \@ifx{#1\undefined}
}%
\providecommand \@ifnum [1]{%
 \ifnum #1\expandafter \@firstoftwo
 \else \expandafter \@secondoftwo
 \fi
}%
\providecommand \@ifx [1]{%
 \ifx #1\expandafter \@firstoftwo
 \else \expandafter \@secondoftwo
 \fi
}%
\providecommand \natexlab [1]{#1}%
\providecommand \enquote  [1]{``#1''}%
\providecommand \bibnamefont  [1]{#1}%
\providecommand \bibfnamefont [1]{#1}%
\providecommand \citenamefont [1]{#1}%
\providecommand \href@noop [0]{\@secondoftwo}%
\providecommand \href [0]{\begingroup \@sanitize@url \@href}%
\providecommand \@href[1]{\@@startlink{#1}\@@href}%
\providecommand \@@href[1]{\endgroup#1\@@endlink}%
\providecommand \@sanitize@url [0]{\catcode `\\12\catcode `\$12\catcode
  `\&12\catcode `\#12\catcode `\^12\catcode `\_12\catcode `\%12\relax}%
\providecommand \@@startlink[1]{}%
\providecommand \@@endlink[0]{}%
\providecommand \url  [0]{\begingroup\@sanitize@url \@url }%
\providecommand \@url [1]{\endgroup\@href {#1}{\urlprefix }}%
\providecommand \urlprefix  [0]{URL }%
\providecommand \Eprint [0]{\href }%
\@ifxundefined \urlstyle {%
  \providecommand \doi  [0]{\begingroup \@sanitize@url \@doi}%
  \providecommand \@doi [1]{\endgroup \@@startlink {\doibase
  #1}doi:\discretionary {}{}{}#1\@@endlink }%
}{%
  \providecommand \doi  [0]{doi:\discretionary{}{}{}\begingroup
  \urlstyle{rm}\Url }%
}%
\providecommand \doibase [0]{http://dx.doi.org/}%
\providecommand \Doi [0]{\begingroup \@sanitize@url \@Doi }%
\providecommand \@Doi  [1]{\endgroup\@@startlink{\doibase#1}\@@Doi}%
\providecommand \@@Doi [1]{#1\@@endlink}%
\providecommand \selectlanguage [0]{\@gobble}%
\providecommand \bibinfo  [0]{\@secondoftwo}%
\providecommand \bibfield  [0]{\@secondoftwo}%
\providecommand \translation [1]{[#1]}%
\providecommand \BibitemOpen [0]{}%
\providecommand \bibitemStop [0]{}%
\providecommand \bibitemNoStop [0]{.\EOS\space}%
\providecommand \EOS [0]{\spacefactor3000\relax}%
\providecommand \BibitemShut  [1]{\csname bibitem#1\endcsname}%
%</preamble>
\end{thebibliography}%


\begin{thebibliography}{99}
\bibitem{knight95} J.B. Knight et al., Phys. Rev. E {\bf 51} (1995) 3957 .
\bibitem{philippe02} P. Philippe and D. Bideau, Eur. Phys. Lett. {\bf 60}, (2002) 677.
\bibitem{pouliquen03} O. Pouliquen {\it et al.}, Phys. Rev. Lett. {\bf 91}, (2003) 014301.
\bibitem{vanel99} L. Vanel and E. Cl\'ement, Eur. Phys. J. B {\bf 11}, (1999) 525.
\bibitem{clement97} E. Cl\'ement {\it et al}, Proceedings of the IIIrd Intern. Conf. on Powders \& Grains (Balkema, Rotterdam, 1997)
\bibitem{claudin97} P. Claudin and J.-P. Bouchaud, Phys. Rev. Lett. {\bf 78}, (1997) 231.
\bibitem{geminard03} J.-C. G\'eminard, Habilitation \`a Diriger des Recherches,
{\it Quelques propri\'et\'es m\'ecaniques des mat\'eriaux granulaires immerg\'es}, Universit\'e Joseph Fourier - Grenoble I (2003).
\bibitem{chen06} K. Chen, J. Cole, C. Conger, J. Draskovic, M. Lohr, K. Klein, T. Scheidemantel and P. Schiffer, Nature {\bf 442}, 257 (2006).
\bibitem{chen09} K. Chen, A. Harris, J. Draskovic and P. Schiffer, Gran. Matt. {\bf 11}, 237 (2009).
\bibitem{divoux08} T. Divoux, H. Gayvallet and J.-C. G\'eminard, Phys. Rev. Lett. {\bf 101} (2008) 148303.
\bibitem{divoux09} T. Divoux, I. Vassilief, H. Gayvallet and J.-C. G\'eminard, AIP Conference Proceedings (Eds. M. Nakagawa and S. Luding),
6th International Conference on the Micromechanics of Granular Media (Golden, CO, 2009).
\bibitem{divoux09t} T. Divoux, PhD, {\it Bruit et fluctuations dans les \'ecoulements de fluides complexes}, Ecole Normale Sup\'erieure de Lyon (2009).
\bibitem{divoux10} T. Divoux, Papers in Physics {\bf 2}, 020006 (2010).
\bibitem{moseley} H. Moseley, {\it The mechanical principles of engineering and architecture}, Eds. Wiley \& Halstead, New York (1853).
\bibitem{bouasse} H. Bouasse, {\it Statique}, Biblioth\`eque Scientifique de l'Ing\'enieur et du Physicien, pp. 259-261 (1920) .
\bibitem{croll09} J. G.A. Croll, Proc. R. Soc. A  {\bf 465}, 791-807 (2009).
\bibitem{tamburi74} A. J. Tamburi, GSA bulletin, {\bf 85}, 351-356 (1974).
\bibitem{persson10}B. N. J. Persson, J. Phys.: Condens. Matter {\bf 22}, 265004 (2010).
\bibitem{baumberger06} T. Baumberger and C. Caroli, Adv. Phys. 55, 279-348 (2006).
\bibitem{geminard10} J.-C. G\'eminard and E. Bertin, Phys. Rev. E {\bf 82}, 056108 (2010).
\bibitem{numerical_recipes} M. P. Allen and D. J. Tildesley, Computer Simulation of Liquids, Clarendon, Oxford, 1987.
\end{thebibliography}

\end{document}